\documentstyle[11pt,newpasp,twoside,epsf]{article}
\reversemarginpar

\begin{document}
\title{Jets from compact objects}
 \author{H.C. Spruit}
\affil{Max-Planck-Institut f\"ur Astrophysik, Postfach 1523, D-85740 Garching, Germany}

\begin{abstract}
Some topics in the theory of jets are reviewed. These include jet precession,
unconfined jets, the origin of knots, the internal shock model as a unifying theme 
from protostellar jets to Gamma-ray bursts, relations between the Blandford-Znajek and MHD disk-wind models, and jet collimation in magnetic acceleration models.
\end{abstract}

\section{Jet precession and warped disks}
Precession is measured directly in the jets of SS 433, whose direction varies with an approximate 165d period. Indirect evidence is seen in the morphology of the hot spots of AGN jets. An example is as Cyg A, where the radio lobes show `fossil' hot spots, offset from the present (most luminous) hot spot position by rotation over angles of some 10 degrees. This gives an approximate point symmetric appearance to the lobes of.

If the central engine causes the direction of the jet to change with time 
(precession), its path in space at a given time appears curved, like the spray of 
water from a rotating garden sprinkler. At each point along the instantaneous path
of the jet, there is a slight difference between the direction of fluid motion
and the tangent to the jet's path. In many cases, this may be the simplest 
explanation for apparent bending in FRII jets. Alternatives like redirection by clouds in 
the path of the jet have would be called for only if there is supporting evidence
like the dissipation and decollimation that accompanies the 
redirection of supersonic flows by external obstacles (observe this by directing 
the jet from a garden hose at the tiles on your garden path).

If jets are produced by accreting compact objects, their flow direction is plausibly 
determined either by the rotation axis of the disk, or that of the accreting object. 
The rate at which the rotation axis of the compact object can change is limited by
the rate $\dot M/M$ at which its angular momentum can change by accretion.
The disk itself can change direction more rapidly, for example if the angular 
momentum vector of the gas supplied to the disk changes in time. It then takes 
only the viscous time in the disk for this change to propagate to the inner
region where the jet originates. If the jet is caused by the disk itself, as in the 
magnetic wind model, its direction will naturally follow the orientation of the
inner disk. If, on the other hand, the jet is caused by the rotation of the 
compact object, as in the Blandford-Znajek (1977) model (see also Blandford, 1993), one might at first sight expect the jet direction to be given also by the axis of the rotating hole. In this mechanism, however, a disk must be present to supply the magnetic field to the horizon of the hole and extract the rotation energy. Since the mechanism by itself does not produce a highly collimated jet, it is possible that the collimation of a Blandford-Znajek outflow is also provided by the disk. In this case, the direction of the jet could follow the disk axis even though it is powered by the hole.

In either case, we arrive at changes of orientation of a disk as a likely 
explanation for jet precession (cf. van den Heuvel et al. 1982). The most promising cause for such changes proposed so far is an instability due to irradiation of the disk by the
central source. Such irradiation can cause the outer parts of the disk to
develop a radiation-heated atmosphere which drives a wind (Begelman et al.\ 1983). Schandl and Meyer (1994) have shown that the momentum flux in such a wind can cause the disk to become unstable to bending out-of-the plane, i.e. warping. As soon as the disk is warped, the radiation intercepted by one side of the disk is larger than the other, and the wind pressure on that side larger. The net torque on the disk due to the difference
in wind pressure causes the irradiated part of the disk to precess. At the same 
time, the warp propagates radially by viscous diffusion, and grows in amplitude with time. Shadowing of parts of the disk by warps in regions closer in makes the 
nonlinear development of the warps quite complicated. Schandl and Meyer show how 
this irradiation-driven wind instability can explain the precessing tilted disk in
Her X-1. 

A radiation-driven wind is expected to be important in the outer regions 
of a disk, where the Compton temperature corresponding to the incident X-ray spectrum
is of the order of the escape velocity from the disk. Closer to the compact object,
wind losses by this process are small. Pringle (1996) studied the same instability without an irradiation-induced wind, using only the effect of radiation pressure on the disk. Pringle (1997) follows the evolution of such warps to arbitrary tilt angles, 
including the self-shadowing effects, and concludes that the inner regions of AGN 
disks can tilt over more than 90\deg. Apart from a time-dependent jet direction, this means that one should expect little correlation between jet axis and the plane of the host galaxy. 

The equations for the evolution of warps in thin accretion disks have been corrected with respect to previous treatments, and put on a firm mathematical basis by Ogilvie (1999). He also presents a practical scheme for the numerical treatment of evolving warps of arbitrary amplitude.

A final cause for precession could be the momentum carried by a magnetically accelerated jet. As in the case of irradiation- and wind -induced warping, the reaction of the jet thrust on the disk may make the disk unstable if the thrust depends on the disk inclination. This possibility has apparently not been studied much, so far.

\section{Unconfined jets}
The confinement of jets, i.e. mechanisms opposing the widening and decollimation
of the jet by the internal pressure, have played an important role in the early
interpretations of jet observations (e.g. Begelman et al. 1984). It is useful
to keep in mind the simple possibility of {\em unconfined} jets, i.e. purely 
ballistic flows like the jet from a fire hose, however. (This is sometimes called `inertial confinement'). The rate of unconfined sideways expansion due to internal pressure may actually be small enough to explain narrow jets in many cases, especially in relativistic jets. To see this, assume as an example the decollimation by internal pressure of an initially collimated jet. That is, we assume that the central engine provides a collimated jet and we follow how it widens when it is exposed to  an external vacuum. The widening converts the enthalpy of the gas, 
$w=c_{\rm s}^2/(\gamma-1)$ into kinetic energy of expansion, where $c_{\rm s}$ is the sound speed and $\gamma$ the ratio of specific heats (assumed fixed). The expansion  velocity (perpendicular to the jet axis), as seen in the comoving frame, is thus of  the order of the sound speed. If the Lorentz factor of the jet is $\Gamma$, the 
travel time of the jet (between the central engine and the its termination at the hot 
spot, for example, as seen in the comoving frame), is reduced by a factor $\Gamma$. 
The opening angle $\delta$ of this freely expanding jet is thus 
$\delta=c_{\rm s}/(\beta\Gamma c)$. As an example, even when the internal sound speed  is initially as large as half the speed of light, the opening angle of a freely expanding  jet with $\Gamma=10$ is only 6 degrees.

The most collimated AGN jets are the FRII's, for which Lorentz factors of order 10 are 
invoked, and where dissipation along the jet is small compared with the dissipation at the terminal shocks (the hot spots). In these jets, the above argument shows that collimation by an external medium is probably not neccessary on observed lengths scales (VLBI and up), if the central engine itself (the unresolved part) provides enough collimation. It is  not desirable either, since the inevitable dissipation associated with the interaction with a collimating external agent would probably disagree with the observed low emission from the jet. Curvature of FRII jets could be accounted for by a modest precession rate (see section 1).

\section{Knots in jets}
Where resolved, jets in general show knots, i.e. sections of high brightness separated by
low emission intervals. Various mechanisms have been proposed, including internal 
instabilities in the jet, or Kelvin-Helmholtz instability due to interaction with an 
environment. The simplest of all, proposed by Rees (1978) assumes that the central 
engine is not exactly steady, but that the flow speed varies by a modest amount. In
the faster episodes, the flow overtakes the slower bits. If the jet speed is as highly 
supersonic as the observations indicate, this process produces large density contrasts 
and internal shocks . The dissipation in these shocks then produces the observed 
synchrotron emission (in the case of AGN) or molecular hydrogen and atomic emission 
(in the case of protostellar jets). In this model, as opposed to most others, the origin
of the knots observed at large distances lies in the central engine. If this engine 
produces symmetric jets, one would expect the knots also to be symmetric on both sides. Internal instabilities and interaction with an environment do not naturally produce such symmetry. In FRII jets, the Lorentz factors are generally so large that only the approaching jet is clearly observable, so this prediction can not be easily tested. In the galactic superluminal sources known so far, however (e.g. Mirabel and Rodriguez, 1999), both the approaching and receding jet are seen, and the observed symmetry of knots (distorted by the Doppler effect, and thereby yielding allowing determination of orientation and speed of the jet) clearly argues in favor of the modulated-flow model. Convincing evidence is also given protostellar jets, where the knots (Herbig-Haro  objects) are often quite symmetric. An example is the HH212 jet (Zinnecker et al. 1998), shown in figure 1. Though at larger distances the knots in this object are less symmetric, indicating interaction with the environment, the inner regions are beautifully  symmetric and present a clear case for the modulated-flow model.

\begin{figure}
\plotone{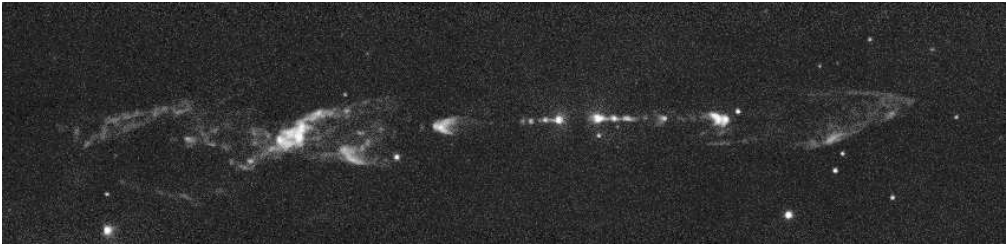}
\caption{The protostellar jet HH212. The symmetry indicates that the knots are formed 
here by modulation of the outflow from the (obscured) source at the center. Courtesy of Thomas Stanke, Mark McCaughrean, and Hans Zinnecker, Astrophysikalisches Institut Potsdam.}
\end{figure}

\subsection{Internal shocks in GRB and Blazars}
Modulation in the central engine also connects naturally with models of  blazars and $\gamma$-ray bursts (GRB) in which internal shocks are invoked, and thereby plays a unifying role for models ranging from the protostellar to the GRB scale. 

In blazars, a systematic variation of the spectral energy distribution with luminosity is observed (Fossati et al. 1998).  In these sources, believed to be AGN with the relativistic jet pointed at the observer, the spectrum is dominated (energy-wise) by two humps, one in the IR-to UV range and one in the gamma range. The first is interpreted as synchrotron emission, the second as Comptonized radiation (both Doppler-shifted by the relativistic motion towards us). The seed photons of the second bump can be either synchrotron radiation generated internally in the jet or external UV radiation from the accretion disk, and are upscattered by the inverse Compton process on energetic electrons in the jet. In order to produce the required energetic electrons, internal shocks are invoked in a model by Ghisellini et al. (1998).  With increasing disk luminosity, the energetic electrons produced by the shocks are cooled more effectively by the inverse Compton process, so that both  the synchrotron and the Compton peak shift to lower photon energy, while the luminosity carried by the Comptonized component increases relative to the synchrotron component. These are the correlations with luminosity noted by Fossati et al. For the model to work, the properties of the jet and the internal shocks are assumed to be rather insensitive to the overall luminosity of the source.

Collimated outflows are also invoked in models for GRB (e.g.\ Meszaros 1999, Sari et al. 1999). In some GRB (in particular the 23 of January 1999 event), the total radiated energy in the $\gamma$ range of can be as high as $3\,10^{54}$ erg, if the source is assumed to radiate isotropically. If the source emits energy only in a narrow cone (pointed at us), the required energy budget can be reduced, bringing currently favored models for the central engine, based on stellar-mass collapsed objects, within the range of relevance. The very erratic light curves of GRB are interpreted, in this class of models, as due to a variation in bulk Lorentz factor of the outflow, caused by a (strong) modulation in the power output of the central engine. Dissipation of the shocks developing in this varying outflow speed accelerates energetic electrons (as in blazar models) producing the observed radiation as synchrotron and synchrotron-self Compton radiation (e.g.  Huang et al. 1998, Chiang and Dermer 1999).

\section{Acceleration mechanisms}
There appear to be 3 acceleration mechanisms for jets still being pursued: the magnetohydrodynamic disk wind, the Blandford-Znajek mechanism, and the `Compton rocket'. The most popular is perhaps the magnetic wind model (Bisnovatyi-Kogan and Ruzmaikin 1976, Blandford 1976, Blandford and Payne 1982, for an introduction and more references see Spruit 1996). This is in part because detailed calculations are possible for this model, resulting in confidence that it is actually realizable in nature. It also has problems that prevent quantitative application to observed systems, however. One of these is the still poorly understood structure of the magnetic field in the disk. A global, ordered poloidal magnetic field is usually assumed (see, however, figure 4 in Blandford and Payne, 1982), but it is not clear if such a field can be maintained in a disk in the presence of a turbulent  diffusion (van Ballegooijen 1989, Lubow et al. 1994). Such a diffusion seems indicated by current numerical simulations of magnetic turbulence in accretion disks (e.g. Hawley \& Stone 1998, Armitage 1998).  Related to this is the `launching problem':  the mass flow in the wind depends sensitively on the transition between disk and wind, which in turn depends on details of the disk stratification and field configuration near the disk surface which are poorly known. Not all disks show jets or outflows, and in the ones that do, the outflows are often sporadic in a way which does not correlate very clearly with the mass accretion rate. An example are the jets in 1915+105, which appear intermittently, possibly correlated with certain transitions in X-ray behavior but not with X-ray flux itself. A third problem is that of collimation.

In the `Compton rocket' model (O'Dell 1981, Kondo 1997, Renauld and Henri 1998), it is assumed that the disk produces, near its surface, a plasma consisting mostly of $e^\pm$ pairs. Pair annihilation in this plasma produces radiation which accelerates the remaining plasma outward. The physics of this model is closely related to the `fireball' models for gamma-ray bursts (Paczy\'nski 1986, Goodman 1986), in which an optically thick, high-temperature pair plasma expands as a relativistic outflow. As in the GRB case, the difficulty with the Compton rocket model is finding a plausible scenario for making the energetic pair plasma. 

In the Blandford-Znajek (1977) mechanism the rotation energy of the accreting black hole is used. A magnetic field in the accretion disk feeds field lines through the horizon of the hole. (In the absence of an external source of field lines, the hole would be unmagnetized). Dragging of inertial frames near the rotating hole twists these field lines, putting stress into them which is released as a Poynting flux away from the hole. In the original form of the model, this Poynting flux is assumed to decay into an $e^\pm$-plasma, in much the same way as is believed to happen in the Crab pulsar wind (e.g. Gallant and Arons 1994, Melatos and Melrose 1996). The escaping relativistic pair wind forms a jet (after suitable collimation, for example by the same disk-maintained magnetic field that magnetizes the hole). 

It is not necessary that the wind be exclusively in the form of a pair plasma. The accretion flow inside the last stable orbit carries baryonic matter into the region where the field line twisting occurs, and this matter may also be accelerated out as a wind. Quantitative models in which this happens have been made by Camenzind (in preparation). If the plasma density remains large enough in the outflow for an MHD approximation to be valid, it will remain a normal (ionized gas) plasma rather than a pair plasma. In these models, there is a gradual transition from a disk-generated magnetic wind, at larger distances from the hole, to a wind powered by field lines dragged around in the hole's rotating gravitational field. This shows that, though it is often assumed that the Blandford-Znajek mechanism will produce a jet consisting of pair plasma, it is equally possible that it will result in a normal plasma outflow.

An energetic argument has been given by Livio et al. (1998) in which the jet-powering capacity of a rotating hole is compared with that of the inner regions of an accretion disk. When a magnetic field is twisted, the energy stored in the field by the twisting torque generally limits the twisted field component to a value not larger than the original untwisted field strength (in some volume-averaged sense). When the field is strained further, the field lines open up, and the twisted (azimuthal) field component as well as the torque {\em decrease} again (Aly 1991, 1994, Lynden-Bell \& Boily 1994). 
Approximating the twisting as if it were all occurring at the horizon $r_{\rm h}$, the maximum torque on the hole is thus of the order $B_{\rm h}^2 r_{\rm h}^3$, where $B_{\rm h}$ the field strength at the horizon, and the rate of energy extraction is $\Omega_{\rm h}B_{\rm h}^2 r_{\rm h}^3$. Since $B_{\rm h}$ is provided by the inner regions of the disk, it is not larger than the field strength $B_{\rm d}$ in these regions. The maximum stress exerted on the disk by a magnetic disk wind is of the order $B^2/2\pi$, hence the total wind torque from the inner regions, with radius $r_d$ and area $\sim \pi r_d^2$, has a maximum of the order $B_{\rm d}^2r_{\rm d}^3$. The maximum rate of energy extraction is then $\Omega_{\rm d}B_{\rm d}^2r_{\rm d}^3$, where $\Omega_{\rm d}$ is the rotation rate of the disk. With these estimates, the energy extraction rate from a hole by a wind is less than the {\em maximum} rate of energy extraction by  a magnetic wind from the disk surrounding it, except when the hole rotates near its maximum rate, in which case the two are comparable. 

Though the hole can have a very large rotational energy available, the rate at which this can be extracted does not exceed the rate at which the disk itself can power a magnetic jet. Note that this conclusion is not an argument against the possible importance of the Blandford-Znajek mechanism. It may well be that there are reasons why the disk itself is not able to reach its possible maximum wind power, while the hole onto which it accretes is happily powering a Blandford-Znajek flow. But the reverse may also be the case. 

\subsection{Observational clue: photon drag}
A relativistic jet accelerated inside a dense radiation field will upscatter these photons to higher energies, as in the blazar models mentioned, but in the process the jet also loses kinetic energy.  If a jet is accelerated to its terminal speed close to a disk (accreting at a known rate), the rate of energy loss in some observed systems would dominate the accretion power (for example, the galactic superluminal source GRS1915+105, see Gliozzi et al. 1999).

This problem is solved if the acceleration can be spread out over a large distance (compared with the inner disk), so that the radiation density is low in the region of largest speed. This is not the case with the Compton rocket mechanism, in which the acceleration is assumed to take place at the disk surface. 

\subsection{Poynting flux}
Magnetic acceleration models can satisfy this requirement elegantly. In a hydromagnetic disk wind, for example, the acceleration is gradual, with most of the acceleration (energetically speaking) taking place away from the disk, near the Alfv\'en radius. The acceleration of a hydromagnetic disk wind is usually described in terms of the centrifugal force acting along the field lines (`bead-on-a-wire'), which illustrates that the acceleration is gradual. Alternatively, the acceleration in this model can be described in terms of energy fluxes. Near the disk, the flux is in the form of a Poynting flux, which gradually gets converted (in part) into a kinetic energy flux. The two descriptions are equivalent. Thus, it is not necessary to appeal exclusively to a Blandford-Znajek process when observations indicate the need for a Poynting flux. An MHD wind will do just as well.

\subsection{$e^\pm$-Winds from accretion disks}
The distinction between the observational consequences of the Blandford-Znajek and the magnetic disk wind mechanisms is further blurred by the possibility that disks may produce a pair dominated wind instead of a normal plasma flow. Supposing that the disk has a strong poloidal magnetic field (i.e. with field lines sticking out above the disk). If conditions are right, disk matter is accelerated up along the field centrifugally, causing an ordinary plasma flow. (The conditions are that the temperature in the disk atmsphere  are high enough and/or the inclination of the field lines with respect to the vertical large enough). 

On the other hand, if these conditions are not satisfied, and mass flow from the disk into the wind region inhibited, the rotating magnetic field of the disk will still produce effects like in pulsar magnetospheres such as that of the Crab pulsar (e.g.  Michel 1991, Gallant and Arons 1994). The enormous electric field strengths associated with the rotating vacuum magnetic field near a black hole accelerates any stray plasma particles to energies sufficient to create pairs. These are themselves accelerated and create a pair cascade until a sufficiently dense pair plasma is produced to limit the electric field by plasma currents. This plasma is then accelerated outward as an MHD wind in much the same way as a normal ionized gas plasma. Except for the different source of rotational energy, the process would be much the same as in the Blandford-Znajek case, including questions such as what collimates the flow.

\section{The magnetic acceleration model}
Details of the magnetic acceleration model have been given on numerous occasions (e.g. Spruit 1996). Only a few of its properties are discussed in the following, in which axisymmetric steady flow assumed. Deviations from axisymmetry and steadyness are probably important in particular for the collimation phase, as discussed below.

The process can be divided conceptually into three stages. In the transition from disk to wind (the `launching' phase), the mass flux into the wind is regulated by the temperature of the plasma and the inclination and strength of the field lines (Blandford and Payne 1982, Ogilvie \& Livio 1998). After passing through the sonic point (located close to the disk surface unless the temperature is near virial) the main acceleration phase sets in, and the wind can be treated as cold to a good approximation (gas pressure negligible). The acceleration is essentially complete at the Alfv\'en surface. Finally, there must be a collimation phase, during and/or after the acceleration phase. 

For non-relativistic flows, the essence of the model is described by the cold (gas pressure neglected) Weber-Davis model (1967, hereafter WD model). This model depends on only one parameter, a magnetization or mass flux parameter. If $\eta=\rho v_{\rm p}/B_{\rm p}$ is the mass flux along a field line, per unit of magnetic flux, then a dimensionless mass flux parameter is
\begin{equation}\mu={4\pi\eta v_{\phi 0}/ B_0}, \end{equation}
where $v_{\phi 0}$ is the orbital velicity of the foot point of the field line at the disk, and $B_0$ the field strength there. The terminal speed of the flow is (Michel, 1969):
\begin{equation} v_\infty=v_{\phi 0}\mu^{-1/3} \end{equation}
At low mass loading $\mu\ll 1$, the final speed exceeds the orbital velocity at the source of the wind. In this case, the centrifugal acceleration picture is a good description of the flow. Inside the Alfv\'en surface, which is far from the disk, the flow corotates approximately with the foot point.  For $\mu\gg 1$, there is no corotating region; instead, the flow is more accurately described as being slowly pushed outward by a highly coiled-up magnetic field.  The transition from low mass flux $\mu<1$ to high mass flux $\mu>1$ has been studied in numerical simulations by Turner et al. (2000). Because of the weak dependence of the terminal velocity on the mass flux, the terminal speed tends to be near the escape speed $v_{\phi 0}$. Large terminal speeds require low mas fluxes, and correspondingly low gas densities in the accelerating region. In numerical simulations, such low densities are hard to deal with because they imply large Alfv\'en speeds that strongly limit the time step. A low-density, high velocity magnetically acccelerated wind is an intrinsically hard numerical problem. 

Qualitative properties of the wind that are not captured by the Weber-Davis model are the collimation (the WD wind is uncollimated) and the asymptotic ratio of Poynting- to kinetic energy fluxes. In an axisymmetric WD wind this ratio is of order unity. This also applies in general if the wind is asymptotically well collimated parallel to the rotation axis. If the poloidal field lines diverge sufficiently rapidly with distance, near the Alfv\'en surface, much more of the Poynting flux is converted into kinetic energy (Begelman and Li, 1994). Such flows are poorly collimated.

If the gas pressure is not neglected, the stationary disk wind (in a Weber-Davis approximation, or if the shape of the poloidal field lines is assumed to be given) depends on two parameters: a mass flux parameter and a temperature parameter. For a concise treatment see Sakurai (1985). 

In the relativistic case the equivalent of the WD model has been given by Michel (1969, 1973). There are now two parameters on which the results depend. In addition to the mass flux parameter, the finite speed of light introduces a relativity parameter $v_\phi/c$, where $v_\phi$ is the rotation velocity of the foot point of a field line. As the mass flux decreases, the Alfv\'en radius asymptotically approaches the light surface (light `cylinder') $r_{\rm L}=rc/v_\phi$ from the inside. Detailed two-dimensional models for steady flows of this kind have been made by Camenzind (1987). For recent time dependent, general-relativistic simulations see Koide et al.\ (1999).

\subsection{Jet (de-)collimation}
In an {\it axisymmetric} steady calculations, excellent collimation of the flow is found in most cases. Many flows become asymptotically parallel to the rotation axis, i.e. the collimation is perfect. The calculations predict that the flow is initially (near the disk surface) poorly collimated. This is because in order for centrifugal acceleration to work, the flow initially has to move away from the axis. Evidence for such poor initial collimation is found in some of the best resolved jets (e.g.\ Junor et al. 1999).
The mechanism that achieves collimation in almost all magnetic models proposed so
 far is the `hoop stress' of the wound-up magnetic field, and this does indeed work 
very well in axisymmetry. It is highly likely, however, that this nearly azimuthal 
field is very unstable to nonaxisymmetric modes which destroy the collimating hoop 
stresses. An external collimating mechanism is probably required for the magnetic 
disk-wind model. In Spruit et al. (1996) it is argued that this ingredient is poloidal 
magnetic flux anchored in the disk at larger distances from the central object.

\end{document}